\documentstyle[12pt,amssymb]{article}
\newcommand{\be}{\begin{equation}}
\newcommand{\ee}{\end{equation}}

\title{{\hfill\small\tt Regular and Chaotic Dynamics\,{\bf 5},\,No.~1,\,89-91
\,(2000)}\\
{\hfill}\\
{A note on geodesics on ellipsoid}\\ 
}
\author{A.M. Perelomov
\footnote{Current e-mail address: perelomo@dftuz.unizar.es}\\
{\footnotesize\em Institute for Theoretical and Experimental
Physics, 117259 Moscow, Russia}}
\date{}

\begin{document}
\maketitle

%\begin{document}
%\title{{\hfill\small\tt Regular and Chaotic Dynamics\,{\bf %5},\,No.~1,\,89-91
%\,(2000)}\\
%{\hfill}\\
%{A note on geodesics on ellipsoid}\\ 
%}
 
\begin{abstract}\noindent
The purpose of this note is to show that the Jacobi problem of geodesics on 
ellipsoid [1], [2], [3] may be reduced to a more simple case, namely, 
to the Clebsch problem [4]. The last one is the problem 
with quadratic nonlinearity and may be solved directly by using of Weber's 
approach [5] in terms of multi-dimensional theta functions.
\end{abstract}

\noindent
{\bf 1.} The problem of geodesics on ellipsoid is a classical one. For 
two-dimensional ellipsoid, its solution was announced by Jacobi on 28 
December, 1838 [1] (see Appendix). Using the remarkable 
substitution, he reduced this problem to quadratures. These results 
were published in the paper [2] and then considered in details with the use 
of elliptic coordinates in his lectures at K\"onigsberg University 
in 1842/43. They can be found  in the book [3]. 

Later  Weierstrass described another approach and succeeded in integrating 
of the equations for geodesics explicitely in terms of two-dimensional 
theta functions [6] .

Another solution of this problem was given by Moser ([7], [8]) 
who included it to the modern scheme of isospectral deformations and 
described the relation of this problem with the geometry of quadrics and 
algebraic geometry of spectral curve. For $n$-dimensional case, the 
algebro-geometrical approach was given by Kn\"orer [9].

Here we show that this problem may be considered without using 
elliptic coordinates and  algebraic geometry as a projection of  a simpler 
system (for a general description of this method, see [10], [11]). 
Namely, it may be considered as a projection 
of the system with a larger number of degrees of freedom and with the 
simplest (quadratic) nonlinearity, the so-called Clebsch system  
[4]. The equations of motion for this system were integrated in terms 
of multi-dimensional theta functions by Weber [5] (for modern 
exposition, see [12]).

\medskip\medskip

\noindent
{\bf 2}. The $(n-1)$-dimensional ellipsoid in $n$-dimensional Euclidean 
space ${\Bbb R}^n$ is defined by the equation
\be Q_0 (x )=\sum_{j=1}^{n}\,\frac{x_j^2}{a_j}=1.\ee
The standard metric in Euclidean space induces the metric on ellipsoid. 
The free motion of a point particle of unit mass on such an ellipsoid 
is described by the Hamiltonian
\be  H=\frac{1}{2}\sum_{j=1}^{n} y_{j}^2\,. \ee
The equations of motion have the form
\be \dot x_j=y_j,\qquad \dot y_j=\ddot x_j=-\,\nu (x,y)\,\frac {x_j}{a_j}\,,
\qquad j=1,\ldots ,n.\ee

To obtain $\nu (x,y)$, we shall differentiate the equation (1) twice with 
respect to  time. As a result, we get
\be \sum _{j=1}^n \frac{x_jy_j}{a_j}=0\ee
and
\be \nu (x,y)=\frac{B(y)}{A(x)},\ee 
where 
\be B(y)=\sum _{j=1}^n \frac {y_j^2}{a_j}\,,
\qquad A(x)=\sum _{j=1}^n \frac {x_j^2}{a_j^2}\,.\ee

\noindent
{\bf Theorem.} {\em Equations} (3), (5), (6) {\em for geodesics on ellipsoid 
are equivalent to the equations of motion for the Clebsch system},
\begin{eqnarray} 
\frac{d}{d\tau }\,y_j&=&-\,\sum _k\omega_{jk}\,y_k, \nonumber \\
&&\\
\frac{d}{d\tau }\,l_{jk} &=&\sum _m (l_{jm}\,\omega _{mk} -\omega _{jm}\,
l_{mk})-\left( a_{j}^{-1} - a_{k}^{-1}\right) y_j\,y_k. \nonumber 
\end{eqnarray}
Here we introduced additional variables related to {\em angular momentum 
tensor}
\be l_{jk}=x_jy_k - x_ky_j\quad \mbox{and}\quad \omega _{jk}=\frac1{a_ja_k}\,
l_{jk} \ee
and $\tau $ is a {\em local time} defined by the formula 
\be d\tau =A^{-1}\,dt,\qquad \frac{d}{d\tau }=A\,\frac{d}{dt}\,.\ee 

\medskip
\noindent
{\bf Proof.} 
After the change of a standard time $t$ by a local time $\tau $, 
the right-hand side of equations (3) takes a polynomial form
\be \frac{d}{d\tau }\,x_j=A(x)\,y_j,\qquad \frac{d}{d\tau }\,y_j=
-\,B(y)\,\frac{x_j}{a_j}\,.\ee

One can see that there are following identities:
\begin{eqnarray} 
B(y)\,\frac{x_j}{a_j}&=&\sum _k \omega_{jk}\,y_k,\qquad 
B(y)=\sum_{j<k}\,\frac{1}{a_j a_k}\,l^2_{jk},\nonumber \\
&&\\
\left( a_{j}^{-1}- a_{k}^{-1}\right) B(y)\,x_jx_k&=&\sum _n (l_{jm}
\omega_{mk}-\omega_{jm}\,l_{mk}) -\left( a_{j}^{-1} - a_{k}^{-1}\right) 
y_jy_k.\nonumber  \end{eqnarray}
From here, it follows equations (7). 
They are the equations with the simplest, namely, quadratic nonlinearity
and they are exactly  equations of motion for the Clebsch system 
[4]. 
\medskip\medskip

\noindent
{\bf 3.} It is easy to check that  equations (7) have $n$ independent 
integrals of motion,
\be 
F_j=y_j^2+\frac{1}{a_j-a_k}\,l_{jk}^{2},\qquad j,k=1,\ldots ,n,\ee
in involution, $\{ F_j, F_k\}=0$. So, the system under consideration 
is completely integrable. 

The generating function of integrals of motion has the form
\[ G_\lambda (y,l)=\sum _{j=1}^n \frac1{a_j-\lambda }\,F_j=\sum _{j=1}^n 
\frac{y_j^2}{a_j-\lambda }-\sum _{j=1}^n \frac{l_{jk}^2}{(a_j-\lambda )
(a_k-\lambda )},\] 
and we have 
\[   \{G_\lambda , G_\mu \}=0. \]

Let us rewrite these equations in the Hamiltonian form. For this, we 
use  the Poisson structure related to the Lie algebra $e(n)$ of motion 
of Euclidean space ${\Bbb R}^n$,
\begin{eqnarray} 
\{ l_{ij}, l_{km}\} &=& \delta_{ik}\,l_{jm} -\delta_{im}\,l_{jk}
-\delta_{jk}l_{im}+\delta_{jm}l_{ik},\nonumber \\
&&\\
\{l_{ij},y_k\}&=&\delta_{jk}\,y_i-\delta _{ik}\,y_j, \qquad \{y_j,y_k\}=0.
\nonumber \end{eqnarray}
So,  the quantities $y_j$, $l_{km}$ generate the Lie algebra $e(n)$ of 
motion in $n$-dimensional Euclidean space.

The Hamiltonian $H$ should be the function of $F_j$ and in fact, due to the 
quadratic form of equations of motion, it should be a linear combination 
of $F_j$, 
\be H=\frac{1}{2}\, b_jF_j .\ee
The simple calculations show that $b_j=a_j^{-1}$. Hence
\begin{eqnarray} 
\dot y_j &=& \{H,y_j\},\qquad   \dot l_{jk}=\{H,l_{jk}\},\nonumber \\
&&\\
H &=& \frac{1}{2}\left( \sum_{j} \frac{y_j^2}{a_j}-\sum _{i,j} 
\frac{l_{ij}^2}{a_ia_j}\right) =G_0(y,l).\nonumber \end{eqnarray}
Eqs. (15) are the Hamiltonian equations related to the Poisson structure for 
the Lie algebra $e(n)$.

For two-dimensional case these equations were integrated explicitly by 
Weber [5] in terms of theta functions of two variables. For 
multi-dimensional case, see [9] and [12]. 

Note that as it was shown by Joachimsthal [13],  the quantity 
\[ I(x,y)=A(x)\,B(y) \]
is an integral of motion, $I(x,y)=c^2$. 

Now we may obtain the expressions for the quantities $x_j$ and $A(x)$:
\be 
x_j(\tau )=-\,B^{-1}(y)\,a_j\,\frac{d}{d\tau }\,y_j,\qquad 
A(x)=\sum _j \frac{x_j^2}{a_j^2}=c^2\,B^{-1}(y).\ee

Finally, the expression for $t$ as the function of $\tau $ is given by 
a quadrature
\be t(\tau )=\int _0^\tau \,A(\tau )\,d\tau .\ee

So, we get the formulae for all dynamical variables.

\newpage

\centerline{\bf Appendix}
\medskip\medskip

Lettre de M. Jacobi \`a M. Arago, concernant les lignes g\'eod\'esiques 
trac\'ees sur un ellipsoide \`a trois axes [1].
\medskip

K\"onigsberg, le 28 d\'ecembre 1838

\,\,\,\, Monsieur,

Je suis parvenu \`a ramener aux quadratures la ligne g\'eod\'esique sur un 
ellipsoide \`a trois axes in\'egaux. Ce sont des formules extr\'emement 
simples, des int\'egrales ab\'eliennes qui se changent dans les int\'egrales 
elliptiques connues, en \'egalant entre eux deux de ces trois axes. Ce 
probleme m'ayant paru long-temps tr\`es difficile, je crois que sa solution 
pourra int\'eresser peut-\`etre quelques-uns des illustres membres de 
l'Acad\'emie des Sciences.
\medskip\medskip

\noindent
{\bf Acknowledgements}
\medskip
I am grateful to  Max--Planck--Institut f\"ur Mathematik in Bonn, 
where this paper was prepared, for the hospitality.


\begin{thebibliography}{999}

\bibitem[1]{Ja1} Jacobi C., {\em Lettre de M. Jacobi \`a M. Arago 
concernant les lignes g\'eod\'esiques trac\'ees sur un ellipsoide \`a trois 
axes}, Compt. Rend. {\bf 8}, 284 (1839)

\bibitem[2]{Ja2} Jacobi C., {\em Note von der geod\"atischen Linie 
   auf einem Ellipsoid und den verschidenen Anwendungen}, JRAM {\bf 19}, 
   309--313 (1839)

\bibitem[3]{Ja3} Jacobi C., in: {\em Vorlesungen \"uber Dynamik} which 
  Jacobi gave at 1842/43 at K\"onigsberg University, Berlin: Reimer (1866)

\bibitem[4]{Cl} Clebsch A., {\em \"Uber die Bewegung eines K\"orpers 
      in einer Fl\"ussigkeit}, Math. Ann. {\bf 3}, 238--262 (1871)
					  
\bibitem[5]{Web} Weber H., {\em Anwendung der Thetafunctionen zweier 
 Ver\"andlicher und die Theorie der Bewegung eines festen K\"orpers in 
 einer Flussigkeit}, Math. Ann. {\bf 14}, 173--206 (1878)

\bibitem[6]{Wei} Weierstrass K., {\em \"Uber die geod\"atischen Linien 
 auf dem dreiachsigen Ellipsoid}, Monatsber. K\"onigl. Akad. Wiss., 
 257--273 (1861)

\bibitem[7]{Mo1} Moser Ju., {\em Various aspects of integrable 
  Hamiltonian systems}, in: {\em Progress in Mathematics} {\bf 8}, 
  {\em Dynamical Systems}, Birkh\"auser, 233--287 (1980)

\bibitem[8]{Mo2} Moser Ju., {\em Geometry of quadrics and spectral 
    theory} in: {\em The Chern Symposium} 1979, Springer--Verlag, 
	147--188 (1980)

\bibitem[9]{Kn} Kn\"orrer H., {\em Geodesics on the ellipsoid}, 
  Invent. Math. {\bf 59}, 119--143 (1980)

\bibitem[10]{OP} Olshanetsky M.A. and Perelomov A.M., {\em Classical 
  integrable finite-dimensional systems related to Lie algebras}, 
  Phys. Reps. {\bf 71}, No.5, 313--400 (1981)

\bibitem[11]{Pe} Perelomov A.M., {\em Integrable Sytems of Classical 
 Mechanics and Lie Algebras}. {\bf I}, Basel: Birkh\"auser (1990)

\bibitem[12]{Du} Dubrovin B.A., {\em Theta functions and nonlinear 
 equations}, Russ. Math. Surv. {\bf 36}, 11--92 (1981)

\bibitem[13]{Jo} Joachimsthal F., {\em Observationes de lineis 
  brevissimis et curvis curvaturae in superficiebus secundi gradus}, 
  JRAM {\bf 26}, 155--171 (1843)

\end{thebibliography}
\end{document}